\documentclass[conference]{IEEEtran} 
\IEEEoverridecommandlockouts
\usepackage{cite}
\usepackage{amsmath,amssymb,amsfonts}
\usepackage{algorithmic}
\usepackage{graphicx}
\usepackage{textcomp}
\usepackage{xcolor}
\def\BibTeX{{\rm B\kern-.05em{\sc i\kern-.025em b}\kern-.08em
    T\kern-.1667em\lower.7ex\hbox{E}\kern-.125emX}}
\begin{document}

\title{Pitch-synchronous DCT features: A pilot study on speaker identification\\
}

\author{\IEEEauthorblockN{Amit Meghanani}
\IEEEauthorblockA{\textit{Department of EE} \\
\textit{Indian Institute of Science}\\
Bangalore, India \\
amitm@iisc.ac.in}
\and
\IEEEauthorblockN{A G Ramakrishnan}
\IEEEauthorblockA{\textit{Department of EE} \\
\textit{Indian Institute of Science}\\
Bangalore, India \\
agr@iisc.ac.in}
}

\maketitle

\begin{abstract}
We propose a new feature, namely, pitch-synchronous discrete cosine transform (PS-DCT), for the task of speaker identification. These features are obtained directly from the voiced segments of the speech signal, without any preemphasis or windowing. The feature vectors are vector quantized, to create one separate codebook for each speaker during training. The performance of the PS-DCT features is shown to be good, and hence it can be used to supplement other features for the speaker identification task. Speaker identification is also performed using Mel-frequency cepstral coefficient (MFCC) features and combined with the proposed features to improve its performance. For this pilot study, 30 speakers (14 female and 16 male) have been picked up randomly from the TIMIT database for the speaker identification task. On this data, both the proposed features and MFCC give an identification accuracy of 90\% and 96.7\% for codebook sizes of 16 and 32, respectively, and the combined features achieve 100\% performance. Apart from the speaker identification task, this work also shows the capability of DCT to capture discriminative information from the speech signal with minimal pre-processing.
\end{abstract}


\section{Introduction}
Speech is produced as a result of the excitation of the time-varying vocal tract system by the time-varying signal through the vocal folds. Assuming the vocal tract configuration to be stationary during the short intervals of time used for analysis, the corresponding speech signal is the convolution of the excitation signal with the vocal tract impulse response in the time domain. For speaker identification, there are time-tested methods to extract information from the voice source (eg. linear prediction residual)[1] 
and the vocal tract filter (eg. MFCC)[2].
The voice source characterization has also been reported earlier using PS-DC[3].

In contrast, in our work, we do not separate the voice source and vocal tract information. Also, we do not utilize any traditional window-sliding approach for feature extraction. We derive the PS-DCT directly from any speech signal for extracting speaker related information. To extract the epochs from the speech signal, we have used the algorithm mentioned in the work[4]. 

A pitch synchronous analysis of voiced sounds was also reported in [5]
\cite{psjasa} 
to show  that vowel sounds can be represented by a sequence of poles arising from the vocal tract and a sequence of zeros characterizing the glottal excitation. Similarly, in 
[6], Zilca et al. have utilized pseudo pitch synchronous analysis of speech for speaker identification task. In this work, the authors have adjusted the presence of pitch information in the speech signal either by interpolating pitch cycles of the LPC residual signal or zero padding the incomplete pitch cycles.

The term 'pitch synchronous' is meaningful only for the voiced part of the speech signal. Since DCT captures both magnitude and phase, it is shift-variant and captures the actual shape of the waveform. Now, in the case of voiced segments, we are taking one complete pitch cycle from peak to peak; hence, the DCT for different pitch cycles for the same voiced phone and for the same speaker are expected to be similar. However, this is not the case for the unvoiced segments, which are aperiodic. In case we use equi-length frames of the unvoiced segments, DCT for different frames for the same unvoiced phone and for the same speaker are not expected to be similar because of its shift-variant property. So, the unvoiced segments need to be treated differently. Hence, only the voiced segments are used in the present work and for selecting the voiced parts of the speech, we have used the labeled information in the TIMIT database. In a practical application, one can make use of any standard algorithm in the literature to detect the voiced segments [7].
Since this is a pilot study involving only thirty speakers, we have chosen a simple vector quantization based representation of each speaker and a nearest-neighbour classification using Euclidean distance. 

Thus, the codebook contains PS-DCT features from the voiced regions. For comparative study, we also use the well-known MFCC features to extract speaker related information. MFCC features mainly contain the vocal tract information of the speaker. Since every speaker is supposed to have different vocal tract structure, these feature are promising for speaker identification task, and have been used widely in the literature. We computed the MFCC features for all the speakers and created VQ codebooks of them too, during training. In this case also, one codebook is created for each speaker.

\begin{figure*}[ht]
  \includegraphics[width=0.8\paperwidth]{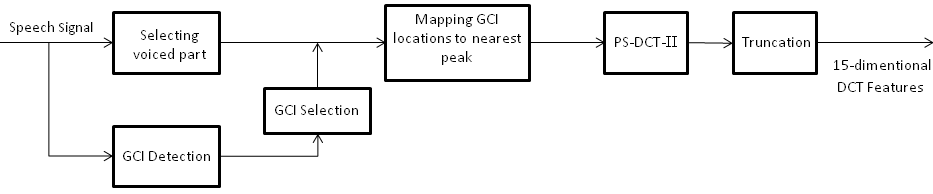}
  \caption{Block diagram for extracting the proposed PS-DCT features from speech. GCIs are detected using any standard algorithm [4].}
  \label{fig:boat1}
\end{figure*}

The motivation for using DCT to derive the spectral features is as follows. In fact, DCT-II of a M-length sequence can be obtained as the first M values of the DFT of a 2M-length, even symmetric sequence, created by an even-symmetric extension of the given signal 
[8]. Thus, DCT captures all the information of a signal that DFT captures. However, most practical applications of DFT consider only the magnitude part of the DFT coefficients and hence, the information contained in the phase is lost. However, since DCT is all real, when we use DCT as features, we do not lose any information contained in the signal. Further, since all the M DCT coefficients obtained are distinct, the frequency resolution of DCT is twice that of DFT. Also, all the basis vectors of DCT start from a peak value and end in a peak or valley. Hence, when the analysis frame is also chosen in such a way that it starts from a peak in the signal to another peak, there is no need to apply a signal shaping window, before applying DCT. This is further justified, since we are only characterizing the signal, and DCT being a linear transformation, is fully reversible. The proposed features contain both the source (vocal fold) and the filter (vocal tract) information. In fact, since the length of the DCT used for analysis is always equal to the pitch period (which varies across speakers), the DCT coefficients correspond to multiples of the pitch frequency. 

\section{Extraction of pitch-synchronous DCT features}
The complete process of feature extraction is shown by the block schematic in Fig. 1.

\subsection{Pre-processing of the speech utterances}

We have used the TIMIT database 
[9]for this work. It has speech data from 630 speakers, sampled at 16 KHz and stored at a  resolution of 16 bits/sample. For our study, we have randomly selected 16 male and 14 female speakers. Out of the ten utterances available from each speaker, we have chosen 8 utterances for the study. Out of the eight utterances, the feature vectors obtained from 6 utterances are used for training and those from 2 other utterances, for testing. This works out to 16 to 18 seconds of training data and 5 to 6 seconds of test data per speaker. For our pilot study, we have chosen the voiced segments of the speech using the labeling provided in the database.

\subsection{PS-DCT features from peak to peak pitch cycles}
For performing PS-DCT, we need to identify the pitch cycles in the speech signal. We have used the algorithm proposed in [4] 
for obtaining the glottal closure instants (GCI) from the voiced regions of the speech signal. 
The analysis interval is chosen starting from the peak nearest to one GCI to the peak nearest to the next GCI. Before applying DCT, we have normalized the frame to unit energy values. Since DCT has excellent energy compaction property, we need not retain all the coefficients. We have excluded the first coefficient as it shows the mean value of the signal hence it is not contributing any relevant information.

By experimenting on the training speech data, as shown in Table I, it is found that the initial 15  coefficients (excluding the first coefficient) have sufficient speaker information and can be used to form the feature vector. The speaker identification performance does not increase if we use more than 15 coefficients. Thus, we have chosen 15 as the optimum choice for the number of coefficients across all the speakers. Therefore, we truncate the PS-DCT features and retain only the initial 15 coefficients. Pre-emphasis is not applied before taking DCT, since, in any case, we are truncating the DCT coefficients and eliminating the higher frequencies. 

Figure 2 shows a randomly selected voiced segment of a speech signal that shows the GCI locations after mapping them to the nearest peaks. Figure 3 illustrates the justification for truncating the DCT to only 15 coefficients. Figure 3(a) shows one cycle of the vowel /ih/ from the training database. Figure 3(b) shows the amplitudes of the DCT coefficients and also the percentage of the energy discarded from the signal as a function of the number of DCT coefficients retained. The figure shows that the energy discarded is pretty little, beyond about 10 coefficients in this case. Figure 3(c) shows that the signal reconstructed from the truncated 15 coefficients captures the essential waveshape of the vowel cycle. Figure 4 shows the plot of the mean energy captured by the truncated DCT coefficients across the entire training data of all the speakers, as a function of the number of DCT coefficients retained.

\begin{table}
        \centering
        \vspace{0.1cm}
        \caption{Speaker identification (SI) performance of the PS-DCT features (in \%) as a function of the number of DCT coefficients included in the feature vector. Codebook size: 32. The results show that retaining 15 coefficients (excluding the first, which is the mean value of the waveform) is the optimum choice, which capture 90.7\% of the energy in the signal, and achieve 96.7\% accuracy. MEC: mean energy captured by the retained coefficients as percentage of the original.}
        \label{tab01}
        \resizebox{0.36\textwidth}{0.09\textheight}{ 
        \begin{tabular}{|c|c|c|} \hline 
            \textbf{No. of coeffs} & \textbf{MEC in \%} & \textbf{SI Accuracy}\\ \hline
            10 &      81.7    &   93.3     \\
            \hline
            \textbf{15} &      90.7    &   96.7 \\
            \hline 
            20  &   93.4 & 96.7\\
            \hline 
            25  &  94.5  & 96.7  \\
            \hline 
            30 &   95.6  & 96.7\\ 
            \hline 
            35 &   96.3   & 96.7\\ 
            \hline 
            40 &    96.9  & 96.7 \\
            \hline 
        \end{tabular}
        }
    \end{table}

\begin{figure}[htbp]
\centerline{\includegraphics[width = \columnwidth, height=4cm]{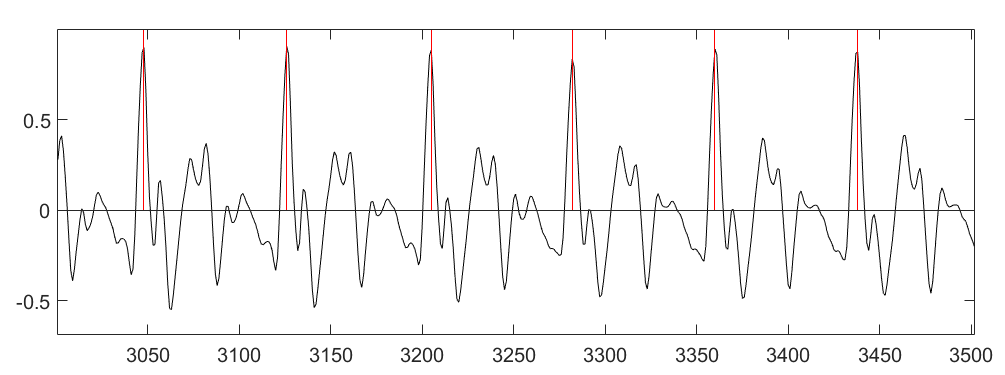}}
\caption{GCI locations in a voiced region of a sample speech signal after mapping them to the nearest peaks. This selection of the analysis frames matches them better to the basis vectors of DCT, all of which start and end in extrema (maxima or minima).}
\label{fig}
\end{figure}

\begin{figure}[htbp]
\centerline{\includegraphics[width=\columnwidth, height=4cm]{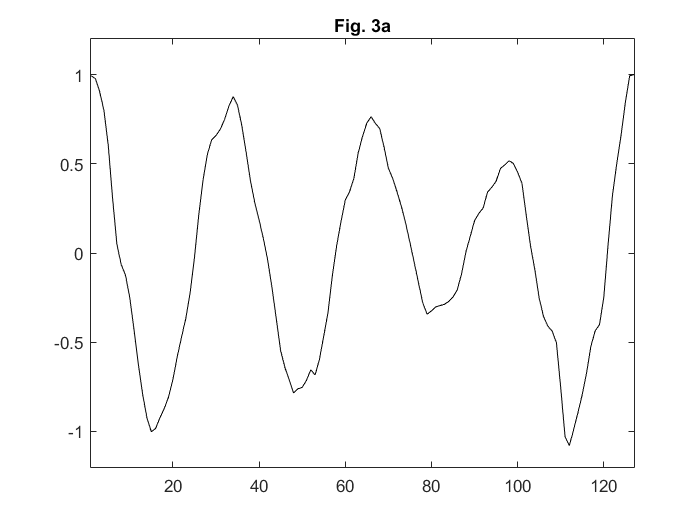}}
\centerline{\includegraphics[width =\columnwidth]{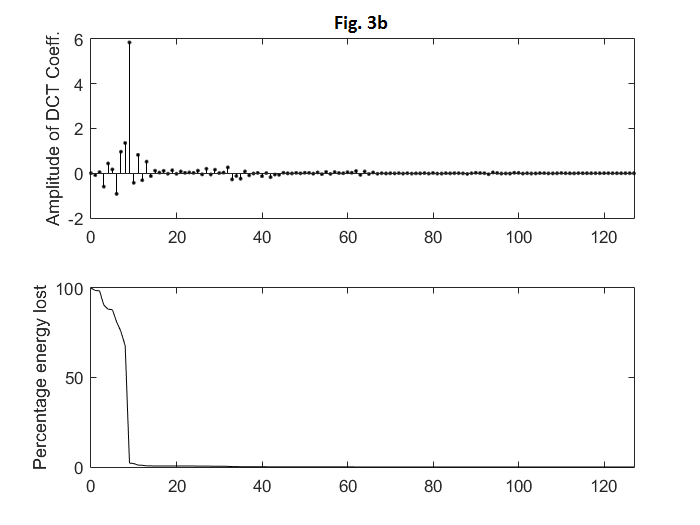}}
\centerline{\includegraphics[width =\columnwidth, height=4cm]{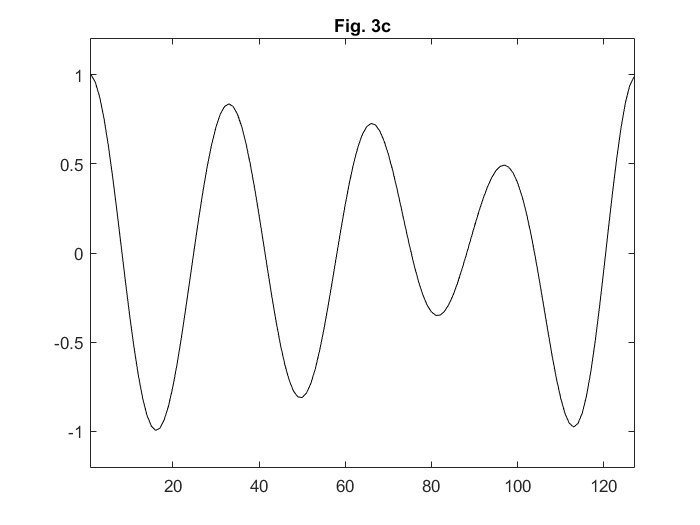}}
\caption{Effective representation of voiced cycles with a limited number of DCT coefficients. (a) The waveform of one pitch period of the vowel /ih/ from peak to peak. (b) The corresponding DCT coefficients (top) and the percentage of the energy in the cycle lost in truncating them (bottom) to different lengths. (c) The vowel waveform, reconstructed from only the first 15 DCT coefficients, captures the essential waveshape.}
\label{fig}
\end{figure}

\begin{figure}[htbp]
\centerline{\includegraphics[width= \columnwidth, height=4.5cm]{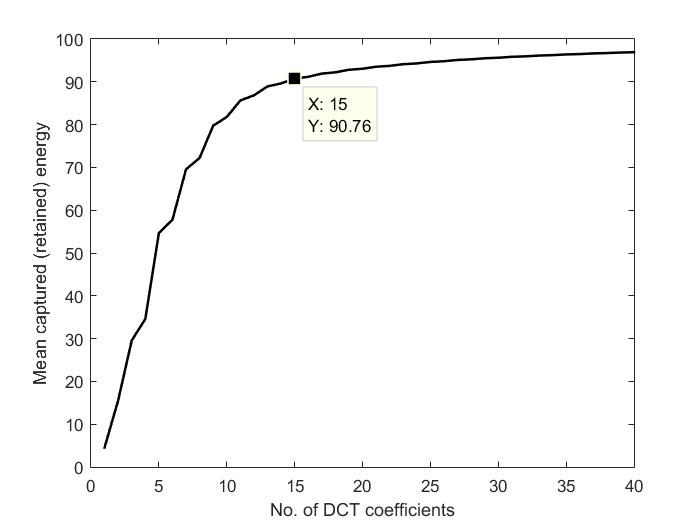}}
\caption{Plot of the mean energy in the pitch period of all the voiced segments from the complete training data of all the 30 speakers chosen from the TIMIT database, as a function of the number of DCT coefficients retained after truncation.}
\label{fig}
\end{figure}

PS-DCT feature vectors are extracted from the six utterances of each speaker, which are used for training. For the male speakers, the number of PS-DCT feature vectors from the training data varied from about 700 to 1740, whereas for the female speakers, it varied from about 1560 to 2580, depending upon their mean pitch value and also the nature of the utterances. This mismatch in the number of training vectors is not an issue, since the pitch cycles are fairly similar for any voiced phone and the same size codebook captures the necessary information, even in the case of female speakers.

\subsection{Generating PS-DCT codebooks for each speaker}
The extracted PS-DCT features are vector quantized. Vector quantization involves non-uniform quantization of the vectors derived from any input signal for compression or pattern recognition [10]
. The feature vectors derived from the entire training data of each speaker are clustered and each cluster is represented by its centroid vector. This collection of all the centroid vectors is known as the codebook and the size of a codebook refers to the number of code vectors. In order to determine the most effective size of the codebook, we generate codebooks of different sizes for each speaker. The data is clustered using k-means clustering algorithm and the distance metric used is Euclidean. The speaker of the codebook with the least minimum distance is identified as the speaker of the test speech.


\section{Results and Discussion}
After creating the codebooks for each speaker with different sizes (16, 32, 64 and 128), we have carried out the testing, using the two test sentences for each speaker.

\subsection{Identification of the test speakers using the VQ codebooks}\label{AA}
Vector quantization of the PS-DCT features extracted from the training data is performed and a codebook is created for each speaker. We have the test data from 30 speakers, where for each speaker, we have 2 utterances as test samples. Feature vectors extracted from all the voiced segments of the two test utterances are used together to make a cumulative decision on the speaker. Suppose k is the size of the codebook, and N is the total number of feature vectors extracted from the test speech signal of a speaker. We compute the Euclidean distance of each test feature vector with each of the k code vectors, independently for each speaker. For a particular speaker, we store these distance values in a Nxk matrix. For each test vector $i$, we find the minimum of the $k$ distances of the i\textsuperscript{th} feature vector from the different code vectors. These values are stored in a Nx1 matrix. Now we compute the sum of all these N values (distances). This sum is the cumulative minimum distance of the test speech from the codebook of a given speaker. Similarly, we compute the cumulative minimum distance (CMD) of the test feature vectors with respect to all the speakers. The speaker of the codebook with the least CMD is identified as the speaker of the test speech.

\subsection{Effect of codebook size on the identification performance}

The performance of the system is evaluated for different choices of the size of the codebook. The results are tabulated in Table II. The identification accuracy is 90\% for a codebook size of 16, since the number of code vectors is nearly the same as the number of the voiced phonemes in American English. The performance increases significantly to 96.7\% for a codebook size of 32 and remains the same as we further increase the size of the codebook beyond 32.

\subsection{Speaker identification using MFCC features and VQ}
The MFCC features are vector quantized as reported in 
[11]. Since MFCC has good vocal tract information, it is widely used as a feature in speaker identification systems. For computing MFCC features, we have used the standard analysis frame lengths of 20 ms with a shift of 10 ms. The MFCCs are computed on the Hanning windowed frames. For every frame, we derive a feature vector of dimension 13. The MFCC features are then vector quantized. Again, we have analyzed the performance of the speaker identification system as mentioned in the previous section and the results are tabulated in Table II. We see that the performance of MFCC features is the same as that of PS-DCT features up to a codebook size of 32, but reaches 100\%, when we increase it further.

\subsection{Convex combination of PS-DCT and MFCC SI systems}
Here we have combined both the speaker identificaton (SI) systems to observe whether PS-DCT has any information that can supplement the MFCC-based system. The performance of the combined system is higher and is 100\% even for a codebook size of 16. From the improved performance of the combined system, we may conclude that the PS-DCT features have some information that supplements the MFCC features and helps the system to perform better. To combine the systems, we use a convex combination of the cumulative minimum distances evaluated from the two systems as follows:
\\

$D\textsubscript{COM} = \alpha D\textsubscript{DCT} +(1 - \alpha)D\textsubscript{MFCC} $  \\

where $\alpha = A\textsubscript{DCT}/(A\textsubscript{DCT} + A\textsubscript{MFCC})$,
\\

where D\textsubscript{DCT}, D\textsubscript{MFCC} and D\textsubscript{COM} are the cumulative minimum  Euclidean distances for the PS-DCT system, MFCC system and the combined system, respectively. A\textsubscript{DCT} and A\textsubscript{MFCC} are the identification accuracies (performance) of the individual PS-DCT and MFCC systems, respectively, obtained in the earlier, separate experiments. The speaker having the  least  combined cumulative minimum distance is identified as the speaker of the test speech. The performance of the combined system is also listed in Table II.

\begin{table}
        \centering
        \vspace{0.1cm}
        \caption{Speaker identification  accuracies (in \%) of PS-DCT, MFCC and combined features on 30 randomly chosen speakers (16 male and 14 female) from the TIMIT database for various codebook sizes. For each speaker, six utterances are used for obtaining the training feature vectors and two, for testing features.}
        \label{tab01}
        \resizebox{0.36\textwidth}{0.09\textheight}{ 
        \begin{tabular}{|c|c|c|c|} \hline 
            \textbf{Codebook size} & \textbf{PS-DCT} & \textbf{MFCC} & \textbf{Combined} \\ \hline 
            16 &  90 & 90 & 100 \\ [5pt]
            \hline 
            32 &  96.7 & 96.7 & 100 \\ [5pt]
            \hline 
            64 &  96.7 & 100 & 100 \\ [5pt]
            \hline 
            128  &  96.7 & 100 & 100 \\ [5pt]
            \hline 
        \end{tabular}
        }
    \end{table}

\subsection{Comparison with other features for speaker identification}
There are other features available for speaker identification studies. For example, LP residual, that mostly contains excitation source information, is utilized for speaker identification in the work by Pati and Prasanna
[11]. In this work, the authors have reported a performance of 73.3\% on TIMIT database for 30 subjects using vector quantization with the codebook size of 256. In contrast, our PS-DCT feature has achieved 96.7\% accuracy on 30 subjects from TIMIT database with a codebook size of 32 only. Thus, PS-DCT seems to be a more promising feature than LP residual.

Another feature, DCTILPR, characterizes the voice source information [3]
. In this work 
[3], using 24 coefficients of DCTILPR, the authors have reported a performance of 94.6\%  on TIMIT database using GMM-based classifier, which is good, considering that the number of test subjects is 168. DCTILPR uses the integrated linear prediction residual, which captures the characteristics of the voice source. In contrast, our PS-DCT feature captures the combined information of voice source and vocal tract with minimal pre-processing. Using only 15 coefficients of PS-DCT and vector quantization, we have achieved 96.7\% accuracy in this pilot study involving only 30 subjects from the TIMIT database.

\section{Conclusion}

We have proposed a computationally simple, but very effective feature for speaker identification, namely, pitch-synchronous discrete cosine transform of the speech signal itself. 
\begin{itemize}
\item 
The results show that PS-DCT features can be derived from the speech signal directly with minimal amount of pre-processing. Even though we do not use any traditional sliding window based feature extraction technique, we are able to effectively extract the speaker related information.
\item
In our work, the analysis is performed directly on the speech signal, similar to MFCC or STFT features. This method extracts both the aspects of the information present in the speech, namely voice source as well as the vocal tract information. The proposed features can possibly supplement other powerful features, such as MFCC or PLP, in various recognition systems to improve the performance.
\item
The results based on this pilot study involving a very limited number of speakers indicate that pitch-synchronous discrete cosine transform has the ability to capture the key information relevant to the identification of the speaker and its efficacy is very close to that of MFCC.
\item
Though the initial experiments are promising, more elaborate experiments are needed to determine its true potential and whether it can scale up to a high number of speakers, by using other classifier, fine tuning the feature extraction process, etc.
\end{itemize}

We have used a single codebook that captures features from only voiced regions.
So unvoiced regions are not dealt yet. However, in the work planned for the future, we intend to use classified vector quantization 
[12], where we use distinct codebooks for the voiced and unvoiced segments for each speaker. A similar approach was used by Pradhan et al.
[13] using the vowel-like and non-vowel like regions separately. However, in our case, we need to come up with an alternative feature for the unvoiced region, which can reliably capture the speaker information in those regions.

\end{document}